\documentstyle[preprint,seceq,mbf]{ptptex}
\notypesetlogo 
 \def\ind{\indent}
 \def\nn{\nonumber}
 \def\be{\begin{equation}}
 \def\ee{\end{equation}}
 \def\ben{\begin{enumerate}}
 \def\een{\end{enumerate}}
 \def\bl{\begin{flushleft}}
 \def\el{\end{flushleft}}
 \def\bt{\begin{tabular}}
 \def\et{\end{tabular}}
 \def\hx{{\hat x}}
 
 \def\hS{{\hat S}}
 \def\hD{{\hat D}}
 \def\hp{{\hat p}}
 \def\ha{{\hat \alpha}}
 \def\hb{{\hat \beta}}
 \def\hP{{\hat P}}
 
 \def\hT{{\hat T}}
 \def\br{\begin{flushright}}
 \def\er{\end{flushright}}
 \def\bc{\begin{center}}
 \def\ec{\end{center}}
 \def\bea{\begin{eqnarray}}
 \def\eea{\end{eqnarray}}
 \def\bea*{\begin{eqnarray*}}
 \def\eea*{\end{eqnarray*}}
 \def\ba{\begin{array}}
 \def\ea{\end{array}}
 \def\bi{\begin{itemize}}
 \def\ei{\end{itemize}}

 \def\cA{{\mbox {${\cal A}$}}}

 \def\bA{{\mbox {{\mbf A}}}}
 
 \def\bB{{\mbox {{\mbf B}}}}
 \def\bC{{\mbox {{\mbf C}}}}
 
 \def\bF{{\mbox {{\mbf F}}}}
 \def\bH{{\mbox {{\mbf H}}}}

 \def\bB{{\mbox {{\mbf B}}}}

 \def\dslash{\partial{\raise 1pt\hbox{$\!\!\!/$}}}

 \def\cham{Chamseddine}
 \def\ind{\indent}
\def\NP{Nucl. Phys.$\;$} 

\def\PL{Phys. Lett.$\;$}

\def\be{\begin{equation}}

 \def\cA{{\mbox {${\cal A}$}}}

 \def\bA{{\mbox {{\mbf A}}}}
 
 \def\bB{{\mbox {{\mbf B}}}}
 \def\bC{{\mbox {{\mbf C}}}}
 \def\bF{{\mbox {{\mbf F}}}}
 \def\bH{{\mbox {{\mbf H}}}}

 \def\bB{{\mbox {{\mbf B}}}}

 \def\vs{\vspace{5mm}}
\title{%
Connes' Gauge Theory on Noncommutative Space-Times
}
\vspace{5mm}
\author{%
  Katsusada {\sc Morita}
}
\inst{%
{\it  
  Nagoya University, Nagoya 464-8602, Japan}
}
\vspace{5mm}
\vspace{10mm}
\abst{%
Connes' gauge theory
is defined on noncommutative space-times.
It is applied to
formulate
a noncommutative Glashow-Weinberg-Salam (GWS) model 
in the leptonic sector.
It is shown that
the model has two
Higgs doublets
and
the gauge bosons sector after the Higgs mechanism
contains
the massive charged gauge fields,
two massless and two massive
neutral gauge fields.
It is also shown that, in the tree level,
the neutrino couples to one of two `photons',
the electron interacts with both `photons' and
there occurs a nontrivial $\nu_R$-interaction
on noncommutative space-times.
Our noncommutative GWS model
is reduced to the GWS theory
in the commutative limit.
Thus in the neutral gauge bosons sector
there are only one massless
photon and only one $Z^0$ in the commutative limit.}
\begin{document}
\maketitle
\section{Introduction}
Connes' 
reconstruction\cite{1),2),3),4),5)} of the standard model
assumes\cite{1)} the two-sheeted Minkowski space-time $M_4\times Z_2$,
the two sheets being separated by the inverse
of order of the weak scale, 
while the Minkowski space-time $M_4$
is assumed to be continuous.
On the other hand, 
there is a growing attention to a
possibility\cite{6),7),8),9),10)}
that
our present space-time geometry would change
and
the space-time coordinates
become noncommutative at very short distances.
The non-commutativity scale
is fundamentally different from
the weak scale and supposed\cite{7)} to be
of order of the Planck length.
The noncommutative geometry\cite{2)}
provides us with a suitable mathematical framework
to describe such a noncommutative space-time structure.
In this paper we ask ourselves
how the two
different scales appear
in the noncommutative gauge theories
(NCGT)\cite{9),10),11),12),13),14),15),16)}
by extending Connes' gauge theory on $M_4\times Z_2$
in the framework of NCGT.
\\
\ind
On noncommutative space-times
characterized by the
commutation relations for the hermitian coordinate
operators $\hx^\mu$
\be
[\hx^\mu,\hx^\nu]=i\theta^{\mu\nu},\quad
\quad\mu,\nu=0,1,2,3,
\label{eqn:1-1}
\ee
where $\theta^{\mu\nu}$ is a real antisymmetric tensor
commuting with $\hx^\rho$,
the spinor $\psi(x)$
should be regarded
as an operator-valued function $\psi(\hx)$,
which is an element of
an algebra $\cA_x$ of 
functions in $\hx^\mu$ modulo the relations (\ref{eqn:1-1}),
and the partial derivative $\partial_\mu\psi(\hx)$
is to be replaced\cite{17)} by the commutator
$[\hp_\mu,\psi(\hx)]$,
where $\hp_\mu$ is defined by
\be
\hp_\mu=-i\theta_{\mu\nu}\hx^\nu,\qquad \theta_{\mu\nu}\theta^{\nu\lambda}
=\delta_\mu^{\;\,\lambda},\qquad [\hp_\mu,\hx^\nu]=\delta_\mu^{\;\,\nu}.
\label{eqn:1-2}
\ee
Here and hereafter we assume that
the matrix $\theta=(\theta^{\mu\nu})$ is invertible.
\\
\ind
There arise new features in
NCGT apart from its nonlocality.
The most prominent one is that
the noncommutative $U(1)$ has a field strength of
Yang-Mills (YM) type.\cite{9),10),11),13),14),15),16)}
The other is that the
YM action but not the
YM Lagrangian are gauge-invariant.
Similarly, 
if the gauge transformation for
$\psi(\hx)$ is acted upon also from the right,
namely,
$\psi(\hx)\to g(\hx)\psi(\hx)u^{\dag}(\hx)$
provided that the matrix multiplication is consistently calculable,
only the Dirac action
becomes gauge-invariant.
We shall argue that,
if the fermion mass is not
gauge-invariant,
the combination of the left and right actions
determines the pattern of the Higgs mechanism
generating the input fermion mass,
yielding a different scale
from that determining the commutation
relations (\ref{eqn:1-1}).
\\
\ind
Connes' interpretation\cite{3)}
of the standard model
regards the Hilbert space of spinors
and their charge conjugates
as a module over the algebra $\cA\otimes \cA^o$,
$\cA^o$ being the opposite algebra of the color-flavor algebra $\cA$.
This essentially means a factorization
of the gauge transformation
for the doubled spinor\cite{18)}
in such a way that
each factor contains flavor and color, separately,
while an Abelian factor is present in both.
The unitary group of the algebra $\cA$
has two $U(1)$s, whereas
the standard model gauge group possesses only one.
This leads to one additional requirement, the unimodularity 
condition\cite{2),3)},
to reconstruct
the standard model in Connes' scheme.
As we have shown recently\cite{19)},
it
happens to determine the correct hypercharge assignment uniquely
if $\nu_R$ exist in each generation.
In this paper, considering the leptonic sector only,
we shall show that
the factorization is naturally obtained by the {\it two-sided}
gauge transformation without introducing the doubled spinor.
\\
\ind
In the next section we define Connes' YM on noncommutative space-times
in the operator formalism and apply it
to
formulate a
noncommutative Glashow-Weinberg-Salam (GWS) model in the leptonic sector, 
which contains two Higgs doublets.
In order to study the Higgs mechanism 
in our noncommutative GWS model,
we rewrite the noncommutative
Connes' YM in terms of the
Weyl-Moyal description\cite{20),21)}
in \S4.
It turns out that
the model
contains 
two massless and two massive
neutral gauge fields in addition to the charged
ones in the gauge bosons sector.
The neutral components become a single massless and a single 
massive neutral gauge fields in the commutative limit
\footnote{
By the commutative limit we always mean the limit $\theta^{\mu\nu}\to 0$
in the Lagrangian level.}.
Similarly the two Higgs doublets become related, leaving
a single standard Higgs doublet,
in the commutative limit.
The final section is devoted to
discussions.
There are two technical Appendices.
\section{Noncommutative Dirac-Yukawa action and noncommutative Connes' YM}%
The free Dirac action reads
\be
\hS_D=(2\pi)^2\sqrt{{\rm det}\theta}
{\rm tr}{\bar\psi}(\hx)(i\gamma^\mu[\hp_\mu,\psi(\hx)]-M\psi(\hx))
=\int\!d^4x{\bar\psi}(x)D\psi(x),
\label{eqn:2-1}
\ee
where
$D=D_0-M$,
$D_0=i\dslash\otimes 1_n$,
$1_n$ being the $n$-dimensional unit matrix,
and 
the ($n$-component) spinor $\psi(x)$
is the Weyl symbol of $\psi(\hx)$
defined 
by\cite{22)}
\be
\psi(x)=\displaystyle{\sqrt{{\rm det}\theta}\over (2\pi)^2}
\int\!d^4ke^{ikx}{\rm tr}(\psi(\hx)\hT^{\dag}(k))
\label{eqn:2-2}
\ee
with $\hT(k)=e^{ik_\mu\hx^\mu}$ and
$\hT^{\dag}(k)=\hT(-k)$.
The trace tr is taken
in the 
Hilbert space in which the operators $\hx^\mu$
are represented,
and normalized
\footnote{
We shall prove the trace formula
tr$\hT(k)=[(2\pi)^2/\sqrt{{\rm det}\theta}]\delta^4(k)$ in the Appendix A.}
to give the last equality in Eq.$\,$(\ref{eqn:2-1}).
\\
\ind
We then require the gauge invariance under
the gauge transformation
\be
\left\{
\ba{l}
\psi(\hx)\to ^g\!\!\psi(\hx)=g(\hx)\psi(\hx)u^{\dag}(\hx),\\[2mm]
{\bar\psi}(\hx)\to ^g\!\!{\bar\psi}(\hx)=
u(\hx){\bar\psi}(\hx)g^{\dag}(\hx),
\ea
\right.\;\;g(\hx)\in M_n(\cA_x),\;\;u(\hx)\in M_1(\cA_x),
\label{eqn:2-3}
\ee
with
$
g(\hx)g^{\dag}(\hx)=g^{\dag}(\hx)g(\hx)={\mbf 1}_n$ and $
u(\hx)u^{\dag}(\hx)=u^{\dag}(\hx)u(\hx)={\mbf 1}$,
where ${\mbf 1}_n$
is the $n$-dimensional unit-operator matrix and
$M_n(\cA_x)$ denotes the set of $n$-dimensional square matrices
with elements in the algebra $\cA_x$.
The gauge invariance
demands the replacement of
the derivative $[\hp_\mu,\psi(\hx)]$ in $\hS_D$ with
the covariant derivative, 
\be
[\hp_\mu,\psi(\hx)]\to [\hp_\mu,\psi(\hx)]+A_\mu(\hx)\psi(\hx)-
\psi(\hx)B_\mu(\hx),
\label{eqn:2-4}
\ee
where the noncommutative gauge fields $A_\mu(\hx)$ and
$B_\mu(\hx)$
are assumed to transform like
\begin{eqnarray}
A_\mu(\hx)&\to& ^g\!\!A_\mu(\hx)=g(\hx)A_\mu(\hx)g^{\dag}(\hx)
+g(\hx)[\hp_\mu,g^{\dag}(\hx)],\nn\\[2mm]
B_\mu(\hx)&\to& ^g\!B_\mu(\hx)=u(\hx)B_\mu(\hx)u^{\dag}(\hx)
+u(\hx)[\hp_\mu,u^{\dag}(\hx)],
\label{eqn:2-5}
\end{eqnarray}
or, equivalently, putting $A=i\gamma^\mu A_\mu$, $B=i\gamma^{\mu T} B_\mu$,
$\hD_0=i\gamma^\mu\hp_\mu$ and
$\hD_0^T=i\gamma^{\mu T}\hp_\mu$,
we have
\begin{eqnarray}
A(\hx)&\to& ^g\!\!A(\hx)=g(\hx)A(\hx)g^{\dag}(\hx)
+g(\hx)[\hD_0,g^{\dag}(\hx)],\nn\\[2mm]
B(\hx)&\to& ^g\!B(\hx)=u(\hx)B(\hx)u^{\dag}(\hx)
+u(\hx)[\hD_0^T,u^{\dag}(\hx)].
\label{eqn:2-6}
\end{eqnarray}
The gauge-invariant, noncommutative Dirac action 
is thus obtained as
\be
\hS_{D+A-B}=(2\pi)^2\sqrt{{\rm det}\theta}
{\rm tr}{\bar\psi}(\hx)(i\gamma^\mu[\hp_\mu,\psi(\hx)]
+A(\hx)\psi(\hx)-\psi(\hx)B(\hx)-M\psi(\hx)),
\label{eqn:2-7}
\ee
where we have assumed that $M$ is gauge-invariant.
\\
\ind
Since $\hp_\mu$ is anti-hermitian,
so is $A_\mu(\hx)$,
$A_\mu^{\dag}(\hx)=-A_\mu(\hx)$ and
similarly for $B_\mu(\hx)$,
ensuring the hermiticity of $\hS_{D+A-B}$.
The noncommutative field strengths
\begin{eqnarray}
F_{\mu\nu}(\hx)&=&
[\hp_\mu,A_\nu(\hx)]-[\hp_\nu,A_\mu(\hx)]
+[A_\mu(\hx),A_\nu(\hx)],\nn\\[2mm]
G_{\mu\nu}(\hx)&=&
[\hp_\mu,B_\nu(\hx)]-[\hp_\nu,B_\mu(\hx)]
+[B_\mu(\hx),B_\nu(\hx)],
\label{eqn:2-8}
\end{eqnarray}
are also anti-hermitian.
Since $[\hp_\mu,\hp_\nu]=i\theta_{\mu\nu}$ commutes with $\hx^\rho$,
the field strengths are
gauge-covariant
\begin{eqnarray}
F_{\mu\nu}(\hx)&\to& 
^g\!F_{\mu\nu}(\hx)=g(\hx)F_{\mu\nu}(\hx)g^{\dag}(\hx),
\nn\\[2mm]
G_{\mu\nu}(\hx)&\to& 
^g\!G_{\mu\nu}(\hx)=u(\hx)G_{\mu\nu}(\hx)u^{\dag}(\hx).
\label{eqn:2-9}
\end{eqnarray}
Consequently, the noncommutative Yang-Mills (NCYM) action
is given by
\be
\hS_{YM}=-\displaystyle{1\over 2}
(2\pi)^2\sqrt{{\rm det}\theta}
{\rm Tr}\displaystyle{1\over g^2}F_{\mu\nu}^{\dag}(\hx)F^{\mu\nu}(\hx)
-\displaystyle{1\over 2g'^2}
(2\pi)^2\sqrt{{\rm det}\theta}
{\rm tr}G_{\mu\nu}^{\dag}(\hx)G^{\mu\nu}(\hx),
\label{eqn:2-10}
\ee
where
Tr includes the trace over
the internal symmetry matrices
in addition to the previously-defined trace tr
and $F^{\mu\nu}(\hx)=g^{\mu\rho}g^{\nu\sigma}F_{\rho\sigma}(\hx)$.
We should delete the second term in the above equation
if the gauge field $B_\mu$ appears already in $F_{\mu\nu}$
in order to avoid the double counting.
\\
\ind
Since the determinant
for the operator-valued gauge function $g(\hx)$
can not  be well-defined,
we can formulate only noncommutative $U(2)$
but not noncommutative $SU(2)$.
(We may extend $2\to N$.)
Moreover, the commutative limit of noncommutative $U(2)$
is $U(1)\times SU(2)$ YM with the same coupling constant.
In order to recover $U(1)\times SU(2)$ YM with the 
different coupling constants
it is preferable to
consider
noncommutative $U(2)$ which is reduced to
$SU(2)$ YM in the commutative limit,
plus additional noncommutative $U(1)^2$
(with the same coupling constant)
reduced to commutative $U(1)$.
In such noncommutative $U(2)$ an Abelian gauge field
mixed with the non-Abelian gauge fields on
noncommutative space-times
would `disappear' in the commutative limit
because it is proportional to $\theta$ for small $\theta$,
while the non-Abelian gauge fields exist for $\theta\to 0$.
If such a model is possible, it will serve to define a 
noncommutative GWS
model which is reduced to
the usual GWS theory in the commutative limit.
We shall argue below that
a noncommutative
Connes' YM may play a role in this direction.
\\
\ind
To define a noncommutative
Connes' YM
we consider\cite{19)} the `gauge' transformations
\be
\left\{
\ba{l}
\psi(\hx)\to b_i(\hx)\psi(\hx)c_i^{\dag}(\hx),\\[2mm]
{\bar\psi}(\hx)\to d_i^{\dag}(\hx){\bar\psi}(\hx)a_i(\hx),
\ea
\right.\;\;a_i(\hx), b_i(\hx) \in M_n(\cA_x),
\;\;c_i(\hx), d_i(\hx) \in M_1(\cA_x)
\label{eqn:2-11}
\ee
with
\be
\displaystyle{\sum_ia_i(\hx)b_i(\hx)}={\mbf 1}_n,\;\;\;
\displaystyle{\sum_ic_i^{\dag}(\hx)d_i^{\dag}(\hx)}={\mbf 1},
\label{eqn:2-12}
\ee
to obtain after taking the sum
over the index $i$ in constructing the sensible action
the gauge fields $A(\hx)$and $B(\hx)$
in Eq.$\;$(\ref{eqn:2-7})
as the sums
\be
A(\hx)=\displaystyle{\sum_ia_i(\hx)[\hD_0,b_i(\hx)]},\;\;\;
B(\hx)=\displaystyle{\sum_ic_i^{\dag}(\hx)[\hD_0^T,d_i^{\dag}(\hx)]}.
\label{eqn:2-13}
\ee
Equation (\ref{eqn:2-13}) is similar to Connes' expression
for YM gauge field.
In fact, in the commutative limit,
we may replace $\hx^\mu\to x^\mu$ and
$\hD_0\to D_0$, obtaining the
noncommutative one-form on $M_4$.
\\
\ind
We define the field strength by the wedge product\cite{19)}
of the Dirac matrices 
\begin{eqnarray}
&&\!\!\!\!\!\! F(\hx)=\displaystyle{\sum_i[\hD_0,a_i(\hx)]\wedge
[\hD_0,b_i(\hx)]}+A(\hx)\wedge A(\hx)
=-\displaystyle{1\over 4}
(\gamma^\mu \gamma^\nu-\gamma^\nu \gamma^\mu)F_{\mu\nu}(\hx),\nn\\[2mm]
&&\!\!\!\!\!\! G(\hx)=\displaystyle{\sum_i[\hD_0^T,c_i^{\dag}(\hx)]\wedge
[\hD_0^T,d_i^{\dag}(\hx)]}+B(\hx)\wedge B(\hx)
=-\displaystyle{1\over 4}
(\gamma^\nu\gamma^\mu 
-\gamma^\mu\gamma^\nu)^TG_{\mu\nu}(\hx),
\label{eqn:2-14}
\end{eqnarray}
where $F_{\mu\nu}(\hx)$ and
$G_{\mu\nu}(\hx)$ are given by
Eq.$\;$(\ref{eqn:2-8}) with
$A_\mu(\hx)=\sum_ia_i(\hx)[\hp_\mu,b_i(\hx)]$
and $B_\mu(\hx)=\sum_ic_i^{\dag}(\hx)[\hp_\mu,d_i^{\dag}(\hx)]$.
NCYM action (\ref{eqn:2-10}) then reads
\be
\hS_{YM}=-\displaystyle{1\over 4g^2}
(2\pi)^2\sqrt{{\rm det}\theta}
{\rm {\mbf Tr}}F(\hx)F(\hx)
-\displaystyle{1\over 4{g'}^2}
(2\pi)^2\sqrt{{\rm det}\theta}
{\rm {\mbf tr}}G(\hx)G(\hx),
\label{eqn:2-15}
\ee
where ${\rm {\mbf Tr}}$ and
${\rm {\mbf tr}}$ includes the trace over the Dirac matrices as well.
The theory defined by the sum $\hS_{D+A-B}+\hS_{YM}$
involves only the physical fields.
\\
\ind
If
$M$ is not gauge-invariant
and fermions exist in chiral multiplets,
we use
the chiral 
decomposition of spinors
so that
the Dirac operator reads
\be
D=D_0+i\gamma_5M, \;D_0=\left(
    \ba{cc}
    i\dslash\otimes 1_{n_L}&0\\
    0&i\dslash\otimes 1_{n_R}\\
    \ea
    \right)\otimes 1_{N_g},\,
M=\left(
  \ba{cc}
  0&M_1\\
  M_1^{\dag}&0\\
  \ea
  \right),
\label{eqn:2-16}
\ee 
with $N_g$ being the number of generations.
The $\gamma_5$ matrix is inserted for later convenience.
The `gauge' transformations (\ref{eqn:2-11})
except for $c_i^{\dag}(\hx)$ and $d_i^{\dag}(\hx)$
are to be extended to those of $2\times 2$ matrices
in the chiral space
\be
f_i(\hx)=\left(
           \ba{cc}
           f_i^L(\hx)&0\\
           0&f_i^R(\hx)\\
           \ea
           \right)\otimes 1_{N_g},\;
f_i^L(\hx)\in M_{n_L}(\cA_x),\;\;\;
           f_i^R(\hx)\in M_{n_R}(\cA_x),\;f=a,b.
\label{eqn:2-17}
\ee
The same procedure as described for the case of the
gauge-invariant $M$
leads to
the generalized noncommutative gauge field
\be
\bA(\hx)=\displaystyle{\sum_i}a_i(\hx)[\hD,b_i(\hx)]
=A(\hx)+i\gamma_5\Phi(\hx),\;
\Phi(\hx)=\displaystyle{\sum_i}a_i(\hx)[M,b_i(\hx)],
\label{eqn:2-18}
\ee
where $\hD=\hD_0+i\gamma_5M{\mbf 1}$ 
and $A(\hx)=\sum_ia_i(\hx)[\hD_0,b_i(\hx)]
=${\scriptsize$\left(
      \ba{cc}
      A^L(\hx)&0\\
      0&A^R(\hx)\\
      \ea
      \right)$}$\otimes 1_{N_g}$.
      The gauge field $B(\hx)$
remains the same as before.
The fields $\bA(\hx)$ and $B(\hx)$ 
appear in the noncommutative Dirac-Yukawa action
\begin{eqnarray}
\hS_{\mbf D}&=&(2\pi)^2\sqrt{{\rm det}\theta}
{\rm tr}{\bar\psi}(\hx)(i\gamma^\mu[\hp_\mu,\psi(\hx)]
+\bA(\hx)\psi(\hx)-\psi(\hx)B(\hx)+i\gamma_5M\psi(\hx))\nn\\[2mm]
&=&(2\pi)^2\sqrt{{\rm det}\theta}
{\rm tr}{\bar\psi}(\hx)(i\gamma^\mu[\hp_\mu,\psi(\hx)]
+A(\hx)\psi(\hx)-\psi(\hx)B(\hx)+i\gamma_5H(\hx)\psi(\hx))
\label{eqn:2-19}
\end{eqnarray}
with $H(\hx)=\Phi(\hx)+M{\mbf 1}$.
\\
\ind
The gauge transformation
\be
\bA(\hx)\to ^g\!\!\bA(\hx)=g(\hx)\bA(\hx)g^{\dag}(\hx)
+g(\hx)[\hD,g^{\dag}(\hx)]
\label{eqn:2-20}
\ee
is induced by
$b_i(\hx)\to b_i(\hx)g^{\dag}(\hx)$ and $a_i(\hx)\to g(\hx)a_i(\hx)$,
where
\be
g(\hx)=\left(
           \ba{cc}
           g_L(\hx)&0\\
           0&g_R(\hx)\\
           \ea
           \right)
           \otimes 1_{N_g},
           \;\;\;g_L(\hx)\in M_{n_L}(\cA_x),\;\;\;g_R(\hx)\in 
M_{n_R}(\cA_x),
\label{eqn:2-21}
\ee
with the conditions $g_L(\hx)g_L^{\dag}(\hx)=g_L^{\dag}(\hx)g_L(\hx)
={\mbf 1}_{n_L}$ and $g_R(\hx)g_R^{\dag}(\hx)=g_R^{\dag}(\hx)g_R(\hx)
={\mbf 1}_{n_R}$. \\
\ind
In order to construct the bosonic
action we again employ the wedge product\cite{19)} of the Dirac matrices
to define the
generalized noncommutative field strength
\be
\bF\,(\hx)=\displaystyle{\sum_i[\hD,a_i(\hx)]\wedge
[\hD,b_i(\hx)]}+\bA(\hx)\wedge \bA(\hx)=
F(\hx)-i\gamma_5[\hP,H(\hx)]-1_4\otimes Y_0(\hx),
\label{eqn:2-22}
\ee
where $\hP=i\gamma^\mu \hP_\mu$ with $\hP_\mu=\hp_\mu+A_\mu$, and
\be
Y_0(\hx)=H^2(\hx)-M^2+y(\hx),\;\;\;y(\hx)\equiv
-\displaystyle{\sum_i}a_i(\hx)[M^2,b_i(\hx)].
\label{eqn:2-23}
\ee
\ind
Unfortunately, however,
there is a nuisance in this definition
because $\bF\,(\hx)$ does not
vanish even when $\bA(\hx)=\sum_ia_i(\hx)[\hD,b_i(\hx)]=0$.
This is a common feature\cite{1),2),3),4),5)} in Connes' YM,
which arises from the ambiguity in defining
the exterior derivative as given by the first term
in Eq.$\,$(\ref{eqn:2-22}) based on the sum (\ref{eqn:2-18}).
\\
\ind
To overcome the difficulty
we resort to
a subtraction method similar to
Connes' one\cite{1),2),3),4),5)} of introducing a quotient algebra.
It consists of
subtracting off
the piece $\langle \bF\,(\hx)\rangle$,
which 
is a matrix of the {\it same form}
\footnote{
For instance, a matrix {\scriptsize$\left(
                                    \ba{cc}
                                    A&B\\
                                    C&D\\
                                    \ea
                                    \right)$}
is of the same form as  {\scriptsize$\left(
                                    \ba{cc}
                                    A'&B'\\
                                    C'&D'\\
                                    \ea
                                    \right)$} if both
                                    are hermitian.
The subtracted
piece is uniquely determined by the orthogonality.}
as
$\sum_i[\hD,a_i(\hx)]\wedge[\hD,b_i(\hx)]$ with
$\bA(\hx)=\sum_ia_i(\hx)[\hD,b_i(\hx)]=0$,
from $\bF\,(\hx)$.
The genuine noncommutative generalized field strength 
is then given by
$[\bF\,(\hx)]=\bF\,(\hx)-\langle \bF\,(\hx)\rangle$.
Since $\sum_i[\hD,a_i(\hx)]\wedge[\hD,b_i(\hx)]|_{{\mbf A}(\hx)=0}
=-1_4\otimes y(\hx)$,
we have $\langle \bF\,(\hx)\rangle=-1_4\otimes \langle Y_0(\hx)\rangle$
where $\langle Y_0(\hx)\rangle$ is a matrix of the {\it same form}
as
$y(\hx)$.
Consequently, we obtain
\be
[\bF(\hx)]=F(\hx)-i\gamma_5[\hP,H(\hx)]-1_4\otimes [Y_0(\hx)],\;\;\;
[Y_0(\hx)]=Y_0(\hx)-\langle Y_0(\hx)\rangle,
\label{eqn:2-24}
\ee
leading to the noncommutative Yang-Mills-Higgs
(NCYMH) action
\begin{eqnarray}
\hS_{YMH}&=&-\displaystyle{1\over 4N_g}
(2\pi)^2\sqrt{{\rm det}\theta}\,
{\rm {\mbf Tr}}_{cg}\displaystyle{1\over g^2}[\bF\,(\hx)][\bF\,(\hx)]
-\displaystyle{1\over 4{g'}^2}
(2\pi)^2\sqrt{{\rm det}\theta}
{\rm {\mbf tr}}G(\hx)G(\hx)\nn\\[2mm]
&=&\hS_{YM}+\displaystyle{1\over N_g}
(2\pi)^2\sqrt{{\rm det}\theta}
{\rm Tr}_{cg}\displaystyle{1\over 
g^2}[\hP_\mu,H(\hx)][\hP^\mu,H(\hx)]\nn\\[2mm]
&&-\displaystyle{1\over 2N_g}
(2\pi)^2\sqrt{{\rm det}\theta}
{\rm Tr}_{cg}\displaystyle{1\over g^2}[Y_0(\hx)]^2,
\label{eqn:2-25}
\end{eqnarray}
where the subscripts $c$ and $g$ of ${\mbf Tr}_{cg}$
and Tr$_{cg}$
indicate the traces in the chiral and generation spaces,
respectively,
and $\hP^\mu=g^{\mu\nu}\hP_\nu$.
\\
\ind
It is necessary to fix the model
in order to make the subtraction
$[Y_0(\hx)]=Y_0(\hx)-\langle Y_0(\hx)\rangle$.
A noncommutative
GWS model in the leptonic sector
is obtained by
taking $n_L=n_R=2$ with
$M_{n_L=2}(\cA_x)\to\bH(\cA_x)$ and $M_{n_R=2}(\cA_x)\to
\bB(\cA_x)$, where
\begin{eqnarray}
&&\left(
       \ba{cc}
       \alpha(\hx)&\beta(\hx)\\
       -\beta^{\dag}(\hx)&\alpha^{\dag}(\hx)\\
       \ea
       \right)\in\bH(\cA_x),\;\;
\left(
       \ba{cc}
       b(\hx)&0\\
       0&b^{\dag}(\hx)\\
       \ea
       \right)\in\bB(\cA_x)
\nn
\end{eqnarray}
with $\alpha(\hx), \beta(\hx),b(\hx)\in M_1(\cA_x).$
In this model
the left-handed fermions are doublets 
like
{\scriptsize$\left(
             \ba{l}
             \nu\\
             e\\
             \ea
             \right)_L$}
and
the right-handed fermions singlets like 
             {\scriptsize$\left(
             \ba{l}
             \nu_R\\
             e_R\\
             \ea
             \right)$} in $N_g$ generations
with the mass matrix
\begin{eqnarray}
M_1&=&\left(
    \ba{cc}
    m_1&0\\
    0&m_2\\
    \ea
    \right), \;\;\;m_{1,2}:\;N_g\times N_g \;{\rm matrices}.
\nn
\end{eqnarray}
\ind
It is then straightforward to show that
\be
H(\hx)=\left(
       \ba{cc}
       0&h(\hx)M_1\\
       M_1^{\dag}h^{\dag}(\hx)&0\\
       \ea
       \right),\;\;h(\hx)=\left(
       \ba{cc}
       {{\phi'_0}}^{\dag}(\hx)&\phi_+(\hx)\\
       -{{\phi'_+}}^{\dag}(\hx)&\phi_0(\hx)\\
       \ea
       \right).
\label{eqn:2-26}
\ee
The two Higgs doublets
\be
\phi(\hx)=\left(
       \ba{l}
       \phi_+(\hx)\\
       \phi_0(\hx)\\
       \ea
       \right),\;\;\;\phi^c(\hx)=\left(
       \ba{l}
       {{\phi'_0}}^{\dag}(\hx)\\
       -{{\phi'_+}}^{\dag}(\hx)\\
       \ea
       \right)
\label{eqn:2-27}
\ee
fuse into a single Higgs doublet in the commutative limit
since the operators defining them become commutative in that limit
\footnote{In the commutative limit $\phi^c(\hx) \to\phi^c(x)$
and $\phi(\hx)\to\phi(x)$ with
$\phi^c(x)=i\sigma_2\phi^{*}(x)$ in terms of the second Pauli
matrix $\sigma_2$.
The change of the spectrum is characteristic to our formulation
of a noncommutative GWS model
which is reduced to the GWS theory in
the commutative limit.}.
It follows from Eq.$\,$(\ref{eqn:2-20}) that,
under the gauge transformation 
by
$g_L(\hx)\in \bH(\cA_x)$ and $g_R(\hx)\in \bB(\cA_x)$
with the conditions
$
g_L(\hx)g_L^{\dag}(\hx)=g_R(\hx)g_R^{\dag}(\hx)={\mbf 1}_2$,
$h(\hx)$ transforms as
\be
h(\hx)\to ^g\!h(\hx)=g_L(\hx)h(\hx)g_R^{\dag}(\hx).
\label{eqn:2-28}
\ee
On the other hand,
the
gauge transformation,
$\psi(\hx)\to
g(\hx)\psi(\hx)u^{\dag}(\hx$),
for the chiral leptons gets factorized 
in the commutative limit
into two factors\cite{18)}
\footnote{
It should be remembered that
the factorization of the
gauge transformations in Connes' scheme
is required to reproduce the correct
hypercharge of leptons using the doubled spinor\cite{18)}
in accord with
Connes' real structure\cite{3)}. 
Here
we do not have to introduce the doubled spinor in order to
obtain the correct charge assignment.}.
\\
\ind 
It can be shown that
$y(\hx)=${\scriptsize$\left(
                      \ba{cc}
                      y_1(\hx)&0\\
                      0&0\\
                      \ea
                      \right)$},
where $y_1(\hx)$ is a hermitian matrix.
On the other hand,
$H^2(\hx)-M^2=${\scriptsize$\left(
                      \ba{cc}
                      y'_1(\hx)&0\\
                      0&y_2(\hx)\\
                      \ea
                      \right)$}, where
$y'_1(\hx)$ is also a hermitian matrix not orthogonal
to
$y_1(\hx)$,
and $y_2(\hx)$
is given by
\begin{eqnarray}
&&y_2(\hx)=\left(
       \ba{cc}
       (\phi^{c\dag}(\hx)\phi^c(\hx)-{\mbf 1})m_1^{\dag}m_1&
       \phi^{c\dag}(\hx)\phi(\hx)m_1^{\dag}m_2\\
       \phi^{\dag}(\hx)\phi^{c}(\hx)m_2^{\dag}m_1&
       (\phi^{\dag}(\hx)\phi(\hx)-{\mbf 1})m_2^{\dag}m_2\\
       \ea
       \right).
\nn
\end{eqnarray}
The result of the subtraction is
$[Y_0(\hx)]=${\scriptsize$\left(
                      \ba{cc}
                      0&0\\
                      0&y_2(\hx)\\
                      \ea
                      \right)$}.
After rescaling NCYMH action reads
\begin{eqnarray}
\hS_{YMH}&=&\hS_{YM}+
\displaystyle{1\over 2}(2\pi)^2\sqrt{{\rm det}\theta}
{\rm Tr}_c[\hD_\mu,h(\hx)]^{\dag}[\hD^\mu,h(\hx)]\nn\\[2mm]
&&-\displaystyle{\lambda'\over 4}
(2\pi)^2\sqrt{{\rm det}\theta}
{\rm tr}[(\phi^{c\dag}(\hx)\phi^c(\hx)-\displaystyle{v^2\over 2}{\mbf 1})^2
{\rm tr}_g(m_1^{\dag}m_1)^2\nn\\[2mm]
&&+\phi^{c\dag}(\hx)\phi(\hx)\phi^{\dag}(\hx)\phi^c(\hx)
{\rm tr}_g(m_1m_1^{\dag}m_2m_2^{\dag})]\nn\\[2mm]
&&-\displaystyle{\lambda'\over 4}(2\pi)^2\sqrt{{\rm det}\theta}
{\rm tr}
[(\phi^{\dag}(\hx)\phi(\hx)-\displaystyle{v^2\over 2}{\mbf 1})^2
{\rm tr}_g(m_2^{\dag}m_2)^2\nn\\[2mm]
&&+
\phi^{\dag}(\hx)\phi^c(\hx)\phi^{c\dag}(\hx)\phi(\hx)
{\rm tr}_g(m_1m_1^{\dag}m_2m_2^{\dag})]
\label{eqn:2-29}
\end{eqnarray}
with $[\hD_\mu,h(\hx)]=[\hp_\mu,h(\hx)]+A_\mu^L(\hx)h(\hx)
-h(\hx)A_\mu^R(\hx)$,
$\hD^\mu=g^{\mu\nu}\hD_\nu$
and tr$_g$ meaning the trace in the generation space.
The parameters $v^2, \lambda'$ are expressed in terms of 
the gauge coupling constants, $N_g$ and the generation-space
traces of the matrices $m_{1,2}$. 
\\
\ind
In NCYMH action (\ref{eqn:2-29})
we are left with only the physical degrees of freedom,
$A_\mu^{L,R}(\hx)$, $B_\mu(\hx)$, $\phi(\hx)$ and $\phi^c(\hx)$.
We now turn to study
the Higgs mechanism
on noncommutative space-times.
\section{Noncommutative GWS model in the leptonic sector}
Since the Higgs mechanism in our
noncommutative GWS model
becomes most transparent in
the Weyl-Moyal description of
the noncommutative Connes' YM,
we shall first translate the operator
language into the function-space language with
deformed product.
\\
\ind
Using the relation
$\hT(k)\hT(k')=e^{{\mbox{\scriptsize${-{i\over 2}k_{1\mu}\theta^{\mu\nu}
k_{2\nu}}$}}}\hT(k+k')$
together with (see Eq.$\,$(\ref{eqn:2-2}))
\begin{eqnarray}
&&\varphi(\hx)=
\displaystyle{1\over (2\pi)^4}\int\!d^4kd^4x\varphi(x)
e^{-ikx}\hT(k),
\nn
\end{eqnarray}
we find\cite{22)} the
basic formulae of the translation
\begin{eqnarray}
\displaystyle{\sqrt{{\rm det}\theta}\over (2\pi)^2}
\int\!d^4ke^{ikx}{\rm tr}(\varphi_1(\hx)\varphi_2(\hx)\hT^{\dag}(k))&=&
\varphi_1(x)*\varphi_2(x),\nn\\[2mm]
\displaystyle{\sqrt{{\rm det}\theta}\over (2\pi)^2}
\int\!d^4ke^{ikx}{\rm 
tr}(\varphi_1(\hx)\varphi_2(\hx)\varphi_3(\hx)\hT^{\dag}(k))
&=&\varphi_1(x)*\varphi_2(x)*\varphi_3(x),
\label{eqn:3-1}
\end{eqnarray}
where the $*$ product is the Moyal product,
\begin{eqnarray}
\varphi_1(x)*\varphi_2(x)&=&
e^{{\mbox{\tiny$\displaystyle{i\over 2}$}}
{\mbox{\tiny$\displaystyle{\partial\over \partial x_1^\mu}$}}
\theta^{\mu\nu}{\mbox{\tiny$\displaystyle{\partial\over \partial 
x_2^\nu}$}}
}\varphi_1(x_1)
\varphi_2(x_2){\big|}_{x_1=x_2=x}.
\nn
\end{eqnarray}
Integration gives
\begin{eqnarray}
(2\pi)^2\sqrt{{\rm det}\theta}
{\rm tr}(\varphi_1(\hx)\varphi_2(\hx))&=&
\int\!d^4x\varphi_1(x)*\varphi_2(x)=\int\!d^4x\varphi_1(x)\varphi_2(x)
,\nn\\[2mm]
(2\pi)^2\sqrt{{\rm det}\theta}
{\rm tr}(\varphi_1(\hx)\varphi_2(\hx)\varphi_3(\hx))
&=&\int\!d^4x\varphi_1(x)*
\varphi_2(x)*\varphi_3(x).
\label{eqn:3-2}
\end{eqnarray}
Using these formulae we rewrite
the `gauge' transformations (\ref{eqn:2-11})
as
\be
\left\{
\ba{l}
\psi(x)\to b_i(x)*\psi(x)*c_i^{\dag}(x),\\[2mm]
{\bar\psi}(x)\to d_i^{\dag}(x)*{\bar\psi}(x)*a_i(x),
\ea
\right.
\label{eqn:3-3}
\ee
where the gauge parameters
$f_i(x)=${\scriptsize$\left(
           \ba{cc}
           f_i^L(x)&0\\
           0&f_i^R(x)\\
           \ea
           \right)\otimes 1_{N_g},
\left\{
\ba{l}
f_i^L(x)\in C^\infty(M_4)\otimes M_{n_L}(\bC)\\
f_i^R(x)\in C^\infty(M_4)\otimes M_{n_R}(\bC)\\
\ea
\right.$}, $f=a,b$
and $c_i(x),d_i(x) \in C^\infty(M_4)\otimes M_1(\bC)$
satisfy $\sum_ia_i(x)*b_i(x)=1_n,n=N_g(n_L+n_R)$ and
$\sum_ic_i^{\dag}(x)*d_i^{\dag}(x)=1$.
The gauge fields are given by
\be
\left\{
\ba{l}
\bA(x)=\displaystyle{\sum_i}a_i(x)*[D,b_i(x)]=A(x)+i\gamma_5\Phi(x),\\
B(x)=\displaystyle{\sum_i}c_i^{\dag}(x)*[D_0^T,d_i^{\dag}(x)],\;\;
D_0^T=i\gamma^{\mu T}\partial_\mu,\\
\ea
\right.
\label{eqn:3-4}
\ee
with
$A(x)=\sum_ia_i(x)*[D_0,b_i(x)]
=${\scriptsize$\left(
      \ba{cc}
      A^L(x)&0\\
      0&A^R(x)\\
      \ea
      \right)$}$\otimes 1_{N_g}$.
The noncommutative Dirac-Yukawa action (\ref{eqn:2-19})
is brought into the form (before rescaling of $H=\Phi(x)+M$)
\begin{eqnarray}
\hS_{\mbf D}&=&\int\!d^4x
{\bar\psi}(x)(D\psi(x)
+*\bA(x)*\psi(x)-*\psi(x)*B(x)+i\gamma_5M\psi(x))\nn\\[2mm]
&=&\int\!d^4x
{\bar\psi}(x)(D_0\psi(x)
+*A(x)*\psi(x)-*\psi(x)*B(x)+i\gamma_5*H(x)*\psi(x)).
\label{eqn:3-5}
\end{eqnarray}
It is gauge-invariant under
\begin{eqnarray}
&&\left\{
\ba{l}
\psi(x)\to g(x)*\psi(x)*U^{\dag}(x),\\[2mm]
{\bar\psi}(x)\to U(x)*{\bar\psi}(x)*g^{\dag}(x),
\ea
\right.\nn\\[2mm]
&&
\left\{
\ba{l}
\bA(x)\to ^g\!\bA(x)=g(x)*\bA(x)*g^{\dag}(x)+g(x)*[D,g^{\dag}(x)],\\[2mm]
B(x)\to ^g\!B(x)=U(x)*B(x)*U^{\dag}(x)+U(x)*[D_0^T,U^{\dag}(x)],
\ea
\right.
\label{eqn:3-6}
\end{eqnarray}
with
\begin{eqnarray}
g(x)&=&\left(
           \ba{cc}
           g_L(x)&0\\
           0&g_R(x)\\
           \ea
           \right)
           \otimes 1_{N_g},\nn\\[2mm]
&&g_L(x)\in C^\infty(M_4)\otimes M_{n_L}(\bC),\;\;\;g_R(\hx)\in 
C^\infty(M_4)\otimes M_{n_R}(\bC),\nn\\[2mm]
&&g(x)*g^{\dag}(x)=1_n,\;\;n=N_g(n_L+n_R),\;\;\;
U(x)*U^{\dag}(x)=1.
\label{eqn:3-7}
\end{eqnarray}
\ind
Let us next turn to the bosonic sector.
The previous model
amounts to replace $M_{n_L=2}(\bC)\to \bH$ and
$M_{n_R=2}(\bC)\to \bB$
\footnote{
Here, $\bH$ is the real quaternions
and $\bB\subset \bH$ is the set of elements {\scriptsize$
                                  \left(
                                  \ba{cc}
                                  b&0\\
                                  0&b^{*}\\
                                  \ea
                                  \right)$} for $b\in \bC$.}.
The YM sector is well-known.
The Higgs kinetic energy term in Eq.$\,$(\ref{eqn:2-29})
is converted into
\be
\hS_{HK}\equiv
\displaystyle{1\over 2}(2\pi)^2\sqrt{{\rm det}\theta}
{\rm Tr}_c[\hD_\mu,h(\hx)]^{\dag}[\hD^\mu,h(\hx)]
=\displaystyle{1\over 2}\int\!d^4x{\rm tr}_c
\{D_\mu,h(x)\}_M^{\dag}*\{D^\mu,h(x)\}_M,
\label{eqn:3-8}
\ee
with
$\{D_\mu,h(x)\}_M=\partial_\mu h(x)+A_\mu^L(x)*h(x)
-h(x)*A_\mu^R(x)$. The tr$_c$ indicates
the trace in the chiral space.
Putting $\hS_{YMH}=\hS_{YM}+\hS_{HK}+\hS_{HP}$
we have the Higgs `potential' term
\begin{eqnarray}
-\hS_{HP}&=&\int\!d^4x\big(
\displaystyle{\lambda'\over 4}
[(\phi^{c\dag}(x)*\phi^c(x)-\displaystyle{v^2\over 2})
*(\phi^{c\dag}(x)*\phi^c(x)-\displaystyle{v^2\over 2})
{\rm tr}_g(m_1^{\dag}m_1)^2\nn\\[2mm]
&&+\phi^{c\dag}(x)*\phi(x)*\phi^{\dag}(x)*\phi^c(x)
{\rm tr}_g(m_1m_1^{\dag}m_2m_2^{\dag})]\nn\\[2mm]
&&+\displaystyle{\lambda'\over 4}
[(\phi^{\dag}(x)*\phi(x)-\displaystyle{v^2\over 2})
*(\phi^{\dag}(x)*\phi(x)-\displaystyle{v^2\over 2})
{\rm tr}_g(m_2^{\dag}m_2)^2\nn\\[2mm]
&&+
\phi^{\dag}(x)*\phi^c(x)*\phi^{c\dag}(x)*\phi(x)
{\rm tr}_g(m_1m_1^{\dag}m_2m_2^{\dag})]\big).
\label{eqn:3-9}
\end{eqnarray}
We find that in the commutative limit the integrand is reduced to
the usual Higgs potential
for a single Higgs doublet.
\\
\ind
The Higgs mechanism occurs if a minimum
of $-\hS_{HP}$ is attained by non-vanishing
vacuum expectation value (VEV)
$\langle\phi(x)\rangle$ of the Higgs field $\phi(x)$.
We seek for the minimum by assuming that
the VEV is constant, $\langle\phi(x)\rangle\equiv
\langle\phi\rangle$,
and $\langle\phi^c\rangle=
i\sigma_2\langle\phi\rangle^{*}$.
In this case
the coefficients of tr$_g(m_1m_1^{\dag}m_2m_2^{\dag})$
vanish
\footnote{
For instance,
$\langle (\phi^{c\dag}(x)*\phi(x)*\phi^{\dag}(x)*\phi^c(x)\rangle
=\langle \phi^{c\dag}\rangle
\langle\phi\rangle
\langle\phi^{\dag}\rangle
\langle\phi^c\rangle
=0$ provided that $\langle\phi^c\rangle=
i\sigma_2\langle\phi\rangle^{*}$.}.
The rest is minimized
if
\be
\langle\phi^{\dag}(x)*\phi(x)\rangle
=\langle\phi^{\dag}\rangle\langle\phi\rangle
=\displaystyle{v^2\over 2},\;\;
\langle\phi^{c\dag}(x)*\phi^c(x)\rangle
=\langle\phi^{c\dag}\rangle\langle\phi^c\rangle
=\displaystyle{v^2\over 2}.
\label{eqn:3-10}
\ee
The gauge transformation for Higgs doublets
is given by
\be
\phi(x)\to ^g\!\!\phi(x)=g_L(x)*\phi(x)*U(x),\;\;\;
\phi^c(x)\to ^g\!\!\phi^c(x)=g_L(x)*\phi(x)^c*U^{\dag}(x),
\label{eqn:3-11}
\ee
where $g_L(x)\in C^\infty(M_4)\otimes \bH$
with $g_L(x)*g_L^{\dag}(x)=1_2$, while
$g_R(x)=${\scriptsize$
                     \left(
                     \ba{cc}
                     U(x)&0\\
                     0&U^{\dag}(x)\\
                     \ea
                     \right)$}$\in C^\infty(M_4)\otimes \bB$
with $U(x)*U^{\dag}(x)=1$.
Remember that
the same function $U(x)$ as in Eq.$\,$(\ref{eqn:3-6})
appears also in $g_R(x)$. Consequently, we
should retain only the first term
in Eq.$\,$(\ref{eqn:2-15}) to define the YM action $\hS_{YM}$.
We assume the unbroken symmetry
\footnote{
This assumption is motivated by generating the input fermion mass
by the Higgs mechanism.}
\begin{eqnarray}
\langle\phi\rangle&\to&
\langle^h\!\phi\rangle
=h_L(x)*\langle\phi\rangle *U(x)
=h_L(x)*U(x)\langle\phi\rangle=\langle\phi\rangle,\nn\\[2mm]
\langle\phi^c\rangle&\to&
\langle^h\!\phi^c\rangle
=h_L(x)*\langle\phi^c\rangle *U^{\dag}(x)
=h_L(x)*U^{\dag}(x)\langle\phi^c\rangle=\langle\phi^c\rangle.
\label{eqn:3-12}
\end{eqnarray}
This together with Eq.$\,$(\ref{eqn:3-10})
has a solution
\begin{eqnarray}
h_L(x)&=&g_R(x)=     \left(
                     \ba{cc}
                     U(x)&0\\
                     0&U^{\dag}(x)\\
                     \ea
                     \right)\in C^\infty(M_4)\otimes \bB,\nn\\[2mm]
&&\langle\phi\rangle=\left(
                      \ba{c}
                      0\\
                      \langle\phi_0\rangle
                      =\displaystyle{v\over \sqrt{2}}\\
                      \ea
                      \right),\;\;
\langle\phi^c\rangle=\left(
                      \ba{c}
                      \langle{\phi'}_0^{\dag}\rangle
                      =\displaystyle{v\over \sqrt{2}}\\
                      0\\
                      \ea
                      \right).
\label{eqn:3-13}
\end{eqnarray}
The unbroken symmetry for leptons is given by
\be
\left\{
\ba{l}
\nu(x)\to U(x)*\nu(x)*U^{\dag}(x),\\
e(x)\to U^{\dag}(x)*e(x)*U^{\dag}(x).\\
\ea
\right.
\label{eqn:3-14}
\ee
\ind
It can be shown that
we are left with two neutral and one charged
massive Higgses among which
only one neutral massive Higgs
to be identified with the standard Higgs remains in the
commutative limit.
\\
\ind
We finally 
investigate the generation of the gauge boson masses.
Remembering Eqs.$\,$(\ref{eqn:3-4}) and (\ref{eqn:3-5})
we put
\footnote{
Both
$A_\mu^L(x)=\sum_ia_i^L(x)*\partial_\mu b_i^L(x)$
and
$A_\mu^R(x)=\sum_ia_i^R(x)*\partial_\mu b_i^R(x)$
are {\it not} traceless in contrast to the model in Ref.18).}
\begin{eqnarray}
A_\mu^L(x)&=&
-\displaystyle{ig\over 2}
\left(
\ba{cc}
A_\mu^0+A_\mu^3&A_\mu^1-iA_\mu^2\\[2mm]
A_\mu^1+iA_\mu^2&A_\mu^0-A_\mu^3\\[2mm]
\ea
\right)\!(x),\nn\\[2mm]
A_\mu^R(x)&=&
-
\displaystyle{ig'\over 2}
\left(
\ba{cc}
B_\mu(x)&0\\[2mm]
0&-C_\mu(x)\\[2mm]
\ea
\right),
\label{eqn:3-15}
\end{eqnarray}
and rescale $B_\mu(x)\to -(ig'/2)B_\mu(x)$ in Eq.$\,$(\ref{eqn:3-5}).
In the commutative limit we have
$A_\mu^0(x)\to 0$ and
$C_\mu(x)\to B_\mu(x)$
\footnote{
The proof will be given in the Appendix B.}.
Namely, the gauge field $A_\mu^L(x)$
is for noncommutative $U(2)$
reduced to commutative $SU(2)$.
Similarly, the gauge field $A_\mu^R(x)$
is for noncommutative $U(1)^2$
(with the same coupling constant)
reduced to commutative $U(1)$.
Consequently, we have two different coupling constants
in the commutative limit as desired for the commutative GWS theory.
Setting
\begin{eqnarray}
h(x)&\to& \langle h\rangle=\displaystyle{v\over \sqrt{2}}
\left(
\ba{cc}
1&0\\
0&1\\
\ea
\right),
\nn
\end{eqnarray}
$\hS_{HK}$ is reduced to the $x$-integral of the mass terms
\begin{eqnarray}
\displaystyle{1\over 2}\int\!d^4x{\rm tr}_c
\{D_\mu,\langle h\rangle\}_M^{\dag}*\{D_\mu,\langle h\rangle\}_M
&=&\int\!d^4x[\displaystyle{1\over 2}M_W^2(W_\mu^{\dag}(x) W^\mu(x)
+W^\mu(x)W_\mu^{\dag}(x))\nn\\[2mm]
&&+\displaystyle{1\over 4}M_Z^2
(Z_\mu(x) Z^\mu(x)+
{Z'}_\mu(x) {Z'}^\mu(x))],
\nn
\end{eqnarray}
where $M_W^2=v^2g^2/4, M_Z^2=v^2(g^2+{g'}^2)/4$ and
\begin{eqnarray}
W_\mu&=&\displaystyle{1\over \sqrt{2}}(A_\mu^1-iA_\mu^2),\nn\\[2mm]
Z_\mu&=&\displaystyle{1\over \sqrt{g^2+{g'}^2}}
        (g(A_\mu^0+A_\mu^3)-g'B_\mu),\nn\\[2mm]
{Z'}_\mu&=&\displaystyle{1\over \sqrt{g^2+{g'}^2}}
        (g(A_\mu^0-A_\mu^3)+g'C_\mu).
\label{eqn:3-16}
\end{eqnarray}
The orthogonal combinations
\begin{eqnarray}
A_\mu&=&\displaystyle{1\over \sqrt{g^2+{g'}^2}}
        (g'(A_\mu^0+A_\mu^3)+gB_\mu),\nn\\[2mm]
{A'}_\mu&=&\displaystyle{1\over \sqrt{g^2+{g'}^2}}
        (g'(A_\mu^0-A_\mu^3)-gC_\mu)
\label{eqn:3-17}
\end{eqnarray}
remain massless, although ${A'}_\mu\to -A_\mu$
in the commutative limit.\\
\ind
The unbroken gauge transformation for mass-eigenstates gauge fields
turns out to be
\begin{eqnarray}
^hW_\mu(x)&=&U(x)*W_\mu(x)*U(x),\nn\\[2mm]
^h\!Z_\mu(x)&=&U(x)*Z_\mu(x)*U^{\dag}(x),\nn\\[2mm]
^h\!{Z'}_\mu(x)&=&U^{\dag}(x)*{Z'}_\mu(x)*U(x),\nn\\[2mm]
^h\!A_\mu(x)&=&U(x)*A_\mu(x)*U^{\dag}(x)+\displaystyle{2i\over e}
           U(x)*\partial_\mu U^{\dag}(x),\nn\\[2mm]
^h\!{A'}_\mu(x)&=&U^{\dag}(x)*{A'}_\mu(x)*U(x)+\displaystyle{2i\over e}
           U^{\dag}(x)*\partial_\mu U(x),
\label{eqn:3-18}
\end{eqnarray}
where we have defined 
\begin{eqnarray}
e&=&\displaystyle{gg'\over \sqrt{g^2+{g'}^2}}.
\nn
\end{eqnarray}
In the commutative limit
we have $A_\mu^0(x)\to 0$ and
$C_\mu(x)\to B_\mu(x)$ so that
${Z'}_\mu(x)\to -Z_\mu(x)$ and ${A'}_\mu(x)\to -A_\mu(x)$, 
the same spectrum as in the neutral gauge bosons sector of the
GWS theory.\\
\ind
We write
the gauge interactions of the chiral fermions as follows:
\begin{eqnarray}
{\bar\psi}(x)*A(x)*\psi(x)&-&{\bar\psi}(x)*\psi(x)*B(x)\nn\\[2mm]
&=&\displaystyle{e\over 2}{\bar\nu}(x)*\gamma^\mu 
(A_\mu(x)*\nu(x)-\nu(x)* A_\mu(x))\nn\\[2mm]
&+&\displaystyle{e\over 2}{\bar e}(x)*\gamma^\mu({A'}_\mu(x)
*e(x)-e(x) *A_\mu(x))\nn\\[2mm]
&+&Z_\mu{\mbox{-}}{\rm interactions}
+{Z'}_\mu{\mbox{-}}{\rm interactions}
+W_\mu{\mbox{-}}{\rm interactions}.
\label{eqn:3-19}
\end{eqnarray}
Looking at $Z_\mu$-interactions for the neutrino
\begin{eqnarray}
\displaystyle{g\over \cos{\theta_W}}
&[&
\displaystyle{1\over 2}(1-\sin^2{\theta_W})
{\bar\nu_L}(x)*\gamma^\mu Z_\mu(x)*\nu_L(x)
-\displaystyle{1\over 2}\sin^2{\theta_W}
{\bar\nu_R}(x)*\gamma^\mu Z_\mu(x)*\nu_R(x)\nn\\[2mm]
&&+\displaystyle{1\over 2}\sin^2{\theta_W}
({\bar\nu_L}(x)*\gamma^\mu \nu_L(x)+
{\bar\nu_R}(x)*\gamma^\mu \nu_R(x))*Z_\mu(x)],
\label{eqn:3-20}
\end{eqnarray}
where the Weinberg angle is defined by
$
\tan{\theta_W}=g'/g$,
we conclude that
$\nu_R$ interacts with $Z_\mu$ 
on noncommutative space-times,
although it escapes the interaction
in the commutative limit as it is gauge-singlet in GWS theory.
In the commutative limit
Eq.$\,$(\ref{eqn:3-14}) is reduced to
{\scriptsize$\left\{
\ba{l}
\nu(x)\to \nu(x),\\
e(x)\to {U^{\dag}}^2(x)e(x),\\
\ea
\right.$}
so that
there is only one photon field $A_\mu=-{A'}_\mu$
and the leptons $(\nu,e)$ have the electric charges $(0, -e)$.
On noncommutative space-times the unbroken symmetry is described
by the gauge transformation (\ref{eqn:3-14}).
Consequently,
in our noncommutative GWS model
in the leptonic sector
there are two `photon' fields, $A_\mu, {A'}_\mu$,
and
two neutral massive gauge fields,
$Z_\mu, {Z'}_\mu$.
It can be seen from E.$\,$(\ref{eqn:3-19})
that, in the tree level,
only one `photon', $A_\mu$, couples to the neutrino, while
both `photons' interact with the electron.
Similarly, the neutrino couples to $Z_\mu$ only
but the electron does to both $Z_\mu$ and ${Z'}_\mu$
in the tree level.
The neutral gauge fields become degenerate into
the photon and $Z^0$, respectively, in the commutative limit.
The structure of $W_\mu$-interactions remain intact.
\section{Discussions}
We have defined Connes' YM on noncommutative space-times.
It contains more physical degrees of freedom
than those in the commutative Connes' YM.
We have considered
a noncommutative GWS model
in the leptonic sector.
The model predicts that, in
addition to the extra massive Higgses,
there are two independent massless
as well as
two independent massive
neutral gauge fields on noncommutative space-times.
They become degenerate into
the photon and $Z^0$, respectively, in the commutative limit.
\\
\ind
In order to include color into the present scheme
we may write
\begin{eqnarray}
l(x)&=&\left(
       \ba{l}
       l_L(x)\\
       l_R(x)\\
       \ea
       \right)\to g(x)*l(x)*U^{\dag}(x),\;\;
       g=\left(
            \ba{cc}
            g_L&0\\
            0&g_R\\
            \ea
            \right),\;\;
            g_R=\left(
                \ba{cc}
                U&0\\
                0&U^{\dag}\\
               \ea
               \right),\nn\\[2mm]
q(x)&=&\Big(
       \left(
       \ba{l}
       q_L^r(x)\\
       q_R^r(x)\\
       \ea
       \right), 
       \left(
       \ba{l}
       q_L^b(x)\\
       q_R^b(x)\\
       \ea
       \right),
       \left(
       \ba{l}
       q_L^g(x)\\
       q_R^g(x)\\
       \ea
       \right)
       \Big)
       \to g(x)*q(x)*v^T(x),
\nn
\end{eqnarray}
where $v(x)\in C^\infty(M_4)\otimes M_3(\bC)$ with
$v(x)*v^{\dag}(x)=v^{\dag}(x)*v(x)=1$.
The new gauge fields associated with $v(x)$
are the gluons.
There is a ninth gluon $G_\mu^0(x)$
which is related to $A_\mu(x)$
via $G_\mu^0(x)=-(1/3)A_\mu(x)$
in the commutative limit
in order to reproduce the correct assignment of the electric
charges of quarks. 
This relation is to be imposed by hand
as opposed to the limit ${A'}_\mu(x)\to -A_\mu(x)$
which is automatic in the leptonic sector.
This may raise a problem
in extending
our noncommutative GWS model 
to a noncommutative standard model. 
This point
will be a subject in a forthcoming paper.
\\
\ind
Non-commutativity of the operator
or Moyal products implies that
a noncommutative generalization of the
conventional field theory model is not unique.
As an example
we consider a
noncommutative QED for leptons $(\nu,e)$
with only a single Abelian gauge field $A_\mu$.
The relevant
gauge transformation is given by
\begin{eqnarray}
\left\{
\ba{l}
\nu(x)\to U(x)*\nu(x)*U^{\dag}(x),\\
e(x)\to U(x)*e(x).
\ea
\right.
\nn
\end{eqnarray}
The gauge couplings are determined as
\begin{eqnarray}
\left\{
\ba{l}
{\bar\nu}(x)*i\gamma^\mu (A_\mu(x)*\nu(x)-\nu(x)*A_\mu(x)),\\
{\bar e}(x)*i\gamma^\mu A_\mu(x)*e(x),
\ea
\right.
\nn
\end{eqnarray}
where the gauge field is assumed to transform like
\begin{eqnarray}
A_\mu(x)\to U(x)*A_\mu(x)*U^{\dag}+U(x)*\partial_\mu U^{\dag}(x).
\nn
\end{eqnarray}
This is inconsistent, however, with the
assumption that
{\scriptsize$\left(
       \ba{l}
       \nu\\
       e\\
       \ea
       \right)_L$}
is a doublet on noncommutative space-times.
In this case both $\nu$
and $e$
should receive the (unbroken) gauge transformation
from both sides, since the neutrino is neutral.
Our gauge transformation (\ref{eqn:3-14})
is chosen to meet this assumption.
But in that case we necessarily have two `photons'
which become a single photon in the commutative limit.
There is a change in the spectrum 
of our noncommutative generalization
of QED for the leptons $(\nu, e)$.
\\
\ind
The non-commutativity parameter 
is very small so that we may work
in the first-order approximation.
We rewrite the $\nu$-$A_\mu$ coupling
in Eq.$\,$(\ref{eqn:3-19})
to the first order in the
non-commutativity parameter as
\begin{eqnarray}
&&-i\displaystyle{e\over 2}\theta^{\rho\sigma}
\partial_\rho
{\bar\nu}(x)
\gamma^\mu\partial_\sigma\nu(x)A_\mu(x),
\nn
\end{eqnarray}
where we have made the partial integration
and used the antisymmetry of $\theta^{\rho\sigma}$.
Similarly, the $\nu_R$-interaction in Eq.$\,$(\ref{eqn:3-20})
is approximated by
\begin{eqnarray}
&&i\displaystyle{g\sin^2{\theta_W}\over 2\cos{\theta_W}}
\theta^{\rho\sigma}
\partial_\rho
{\bar\nu}_R(x)
\gamma^\mu\partial_\sigma\nu_R(x)Z_\mu(x).
\nn
\end{eqnarray}
Next consider the electron-interaction with two `photons'.
We can convert it
to the familiar-looking one
$-e{\bar e}(x)\gamma^\mu e(x)A_\mu(x)$
plus an additional one in the same approximation
\begin{eqnarray}
&&\displaystyle{e\over 2}\theta^{\rho\sigma}
{\bar e}(x)\gamma^\mu e(x)A_{\rho\mu\sigma}(x)
\equiv\displaystyle{e\over 2}{\bar e}(x)\gamma^\mu e(x){\tilde A}_\mu(x),
\nn
\end{eqnarray}
where we put 
${A'}_\mu(x)=-A_\mu(x)+\theta^{\rho\sigma}A_{\rho\mu\sigma}(x).$
Although it is impossible to
cast this extra one into the form
$j^\mu(x) A_\mu(x)$,
we can define the field strength
for ${\tilde A}_\mu(x)=\theta^{\rho\sigma}A_{\rho\mu\sigma}(x)$
by
${\tilde F}_{\mu\nu}(x)=\partial_\mu{\tilde A}_\nu(x)-
\partial_\nu{\tilde A}_\mu(x)+e\theta^{\rho\sigma}
\partial_\rho A_\mu(x)
\partial_\sigma A_\nu(x)$
such that
${F'}_{\mu\nu}(x)=-F_{\mu\nu}(x)+{\tilde F}_{\mu\nu}(x)$
to the first order in $\theta^{\rho\sigma}$
\footnote{To determine the propagator
of ${\tilde A}_\mu$ we should retain a term quadratic
in ${\tilde F}_{\mu\nu}$, which is higher order.
The decomposition
${A'}_\mu(x)=-A_\mu(x)+{\tilde A}_\mu(x)$
defines
${\tilde A}_\mu(x)$
such that,
in the first-order approximation,
the infinitesimal gauge transformation is
$\delta A_\mu=+(2/e)\partial_\mu\alpha-\theta^{\rho\sigma}
\partial_\rho\alpha\partial_\sigma A_\mu$
and $\delta {\tilde A}_\mu=-2\theta^{\rho\sigma}
\partial_\rho\alpha\partial_\sigma A_\mu$,
where $U=(e^{i\alpha})_*=1+i\alpha$.
Consequently, the sum
$-e{\bar e}(x)\gamma^\mu e(x)A_\mu(x)
+(e/2)
{\bar e}(x)\gamma^\mu e(x){\tilde A}_\mu(x)$
upon integration
is gauge-invariant in the same approximation.},
where $F_{\mu\nu}(x)=\partial_\mu A_\nu(x)-
\partial_\nu A_\mu(x)+(e/2)
\theta^{\rho\sigma}
\partial_\rho A_\mu(x)
\partial_\sigma A_\nu(x)$.
\\
\ind
Or, it may be
illegitimate to attempt to
expand a noncommutative GWS model
with respect to the non-commutativity parameter
although the commutative limit
can be discussed already in the Lagrangian level.
We have not yet succeeded in finding an appropriate language
of describing the change of the
spectrum in our theory.
\section*{Acknowledgements}
The author is grateful to
H. Kase, Y. Okumura, S. Kitakado, and H. Ikemori 
for useful discussions.
\vs
\begin{center}
{\large\bf Appendix A}
\end{center}
\vs
In this Appendix we prove the trace formula
tr$\hT(k)=[(2\pi)^2/\sqrt{{\rm det}\theta}]\delta^4(k)$.
The 2-dimensional case was
treated in Ref. 17).\\
\ind
We can always convert the (invertible) matrix $\theta=(\theta^{\mu\nu})$ 
to the canonical form\cite{23)}
$$
\theta=\left(
\ba{cccc}
0&\theta_1&0&0\\
-\theta_1&0&0&0\\
0&0&0&\theta_2\\
0&0&-\theta_2&0\\
\ea
\right),\;\;\;\theta_1\theta_2\not=0.
$$
In this canonical form we have the following commutation relations
$$
[\hx^0,\hx^1]=i\theta_1,\;\;[\hx^2,\hx^3]=i\theta_2,\;\;
{\rm others}=0.
$$
Using the annihilation and creation operators
$\ha=(1/\sqrt{2\theta_1})(\hx^0+i\hx^1), 
\ha^{\dag}=(1/\sqrt{2\theta_1})(\hx^0-i\hx^1),
\hb=(1/\sqrt{2\theta_2})(\hx^2+i\hx^3)$ and 
$\hb^{\dag}=(1/\sqrt{2\theta_2})(\hx^2-i\hx^3)$,
which satisfy $[\ha,\ha^{\dag}]=[\hb,\hb^{\dag}]=1$ and
$[\ha,\hb]=[\ha,\hb^{\dag}]=0$,
we can write
$\hx^0=\sqrt{\theta_1/2}(\ha+\ha^{\dag}),
\hx^1=(1/i)\sqrt{\theta_1/2}(\ha-\ha^{\dag}),
\hx^2=\sqrt{\theta_2/2}(\hb+\hb^{\dag})$ and
$\hx^3=(1/i)\sqrt{\theta_2/2}(\hb-\hb^{\dag})$
so that we have
$e^{ik_\mu\hx^\mu}=e^{A+B+C+D}$, where
$A=\sqrt{\theta_1/2}(ik_0+k_1)\ha,
B=\sqrt{\theta_1/2}(ik_0-k_1)\ha^{\dag},
C=\sqrt{\theta_2/2}(ik_2+k_3)\hb$ and
$D=\sqrt{\theta_2/2}(ik_2-k_3)\hb^{\dag}$.
We next resort to the well-known formula
$$
e^{A+B}=e^Ae^Be^{-{\mbox{\tiny$\displaystyle{1\over 2}$}}[A,B]},
\;\;\;[A,B]:\;c{\rm -number}
$$
to obtain
$$
\hT(k)=
e^{A+B+C+D}=e^Ae^Be^Ce^De^{(\theta_1/4)(k_0^2+k_1^2)+(\theta_2/4)(k_2^2+k_3
^2)}.
\eqno{({\rm A}\cdot 1)}
$$
Since the trace is independent of the basis in the
Hilbert space spanned by $\hx^\mu$,
we evaluate it in the coherent states basis
$$
{\rm tr}\hT(k)=
(\displaystyle{i\over 2\pi})^2\int\!dzdz^{*}d\zeta d\zeta^{*}
\langle z,\zeta|\hT(k)|z,\zeta\rangle e^{-|z|^2-|\zeta|^2},
\eqno{({\rm A}\cdot 2)}
$$
where $|z,\zeta\rangle=e^{z\ha^{\dag}}e^{\zeta\hb^{\dag}}|0\rangle$
with
$\ha|0\rangle=\hb|0\rangle=0$.
Substituting Eq.$\,$(A$\cdot$ 1) into Eq.$\,$(A$\cdot$ 2)
we find 
$$
{\rm tr}\hT(k)=
(\displaystyle{i\over 2\pi})^2\int\!dzdz^{*}d\zeta d\zeta^{*}e^X,
$$
where 
$
X=z^{*}\sqrt{\theta_1/2}(ik_0-k_1)+z\sqrt{\theta_1/2}(ik_0+k_1)
+\zeta^{*}\sqrt{\theta_2/2}(ik_2-k_3)
+\zeta\sqrt{\theta_2/2}(ik_2+k_3)-(\theta_1/4)(k_0^2+k_1^2)
-(\theta_2/4)(k_2^2+k_3^2).
$
Changing the variables by
$z=x^0+ix^1, z^{*}=x^0-ix^1, \zeta=x^2+ix^3, \zeta^{*}=x^2-ix^3$
with
$dzdz^{*}d\zeta d\zeta^{*}=(-2i)^2d^4x$,
we arrive at
\begin{eqnarray*}
{\rm tr}\hT(k)&=&
\displaystyle{1\over \pi^2}\int\!d^4xe^{i[
\sqrt{2\theta_1}(x^0k_0+x^1k_1)+
\sqrt{2\theta_2}(x^2k_2+x^3k_3)]}e^{-(\theta_1/4)(k_0^2+k_1^2)
-(\theta_2/4)(k_2^2+k_3^2)}\\[2mm]
&=&\displaystyle{(2\pi)^4\over \pi^2}
\delta(\sqrt{2\theta_1}k_0)
\delta(\sqrt{2\theta_1}k_1)
\delta(\sqrt{2\theta_2}k_2)
\delta(\sqrt{2\theta_2}k_3)\\[2mm]
&=&\displaystyle{(2\pi)^2\over \theta_1\theta_2}\delta^4(k)
=\displaystyle{(2\pi)^2\over \sqrt{{\rm det}\theta}}\delta^4(k).
\end{eqnarray*}
\vs
\begin{center}
{\large\bf Appendix B}
\end{center}
\vs
Needless to say the gauge fields $A_\mu^{L,R}(x)$ of 
Eq.$\,$(\ref{eqn:3-15})
must become traceless\cite{18)} in the commutative limit.
The purpose of this Appendix is to prove this statement
in our formulation.\\
\ind
By writing
the elements of $C^\infty(M_4)\otimes \bH$ in Eq.$\,$ (\ref{eqn:3-4}) 
as
$a_i^L(x)=${\scriptsize$\left(
       \ba{cc}
       \alpha_i(x)&\beta_i(x)\\
       -\beta_i^{*}(x)&\alpha_i^{*}(x)\\
       \ea
       \right)$} and
       $b_i^L(x)=${\scriptsize$\left(
       \ba{cc}
       \gamma_i(x)&\delta_i(x)\\
       -\delta_i^{*}(x)&\gamma_i^{*}(x)\\
       \ea
       \right)$},
we have
$(A_\mu^L)_{11}(x)=\sum_i(\alpha_i(x)*\partial_\mu \gamma_i(x)
-\beta_i(x)*\partial_\mu \delta_i^{*}(x))$, while
$(A_\mu^L)_{22}(x)=
\sum_i(-\beta_i^{*}(x)*\partial_\mu \delta_i(x)
+\alpha_i^{*}(x)*\partial_\mu \gamma_i^{*}(x))$.
Because of the $*$-product they are independent.
However, in the commutative limit,
we can omit the $*$-symbol so that
$(A_\mu^L)_{11}(x)=\sum_i(\alpha_i(x)\partial_\mu \gamma_i(x)
-\beta_i(x)\partial_\mu \delta_i^{*}(x))
=-\sum_i(\alpha_i^{*}(x)\partial_\mu \gamma_i^{*}(x)
-\beta_i^{*}(x)\partial_\mu \delta_i(x))=
-(A_\mu^L)_{22}(x)$, where we have used the anti-hermiticity.
\\
\ind
On the other hand,
by our choice of $g_R(x)$
$(A_\mu^R)_{11}(x)$ and $B_\mu(x)$
enjoy the same
gauge transformation law so that
we should put $c_i^{*}(x)=(a_i^R)_{11}(x)$ and
$d_i^{*}(x)=(b_i^R)_{11}(x)$,
yielding the equality
$B_\mu(x)=\sum_ic_i^{*}(x)*\partial_\mu d_i^{*}(x)
=\sum_i(a_i^R)_{11}(x)*\partial_\mu (b_i^R)_{11}(x)=(A_\mu^R)_{11}(x)$.
In contrast, $(A_\mu^R)_{22}(x)=\sum_i(a_i^R)_{22}(x)*
\partial_\mu (b_i^R)_{22}(x)$
is not related with $B_\mu(x)$
since $(a_i^R)_{22}(x)=(a_i^{R*})_{11}(x)$ and 
$(b_i^R)_{22}(x)=(b_i^{R*})_{11}(x)$,
and $\sum_i(a_i^{R*})_{11}(x)*\partial_\mu (b_i^{R*})_{11}(x)$
is not equal to 
$-\sum_i(a_i^R)_{11}(x)*\partial_\mu (b_i^R)_{11}(x)=-(A_\mu^R)_{11}(x)$.
As in the previous case, however,
in the commutative limit
we can omit the $*$ symbol and 
$\sum_i(a_i^{R*})_{11}(x)\partial_\mu (b_i^{R*})_{11}(x)=
-\sum_i(a_i^R)_{11}(x)\partial_\mu (b_i^R)_{11}(x)$
by anti-hermiticity.
Hence, $(A_\mu^R)_{11}(x)=-(A_\mu^R)_{22}(x)$
in the commutative limit.

\end{document}